\documentclass[shortnote]{jpsj2} 
\title{Uniform Current in Graphene Strip with Zigzag Edges}

\author{Shuhei \textsc{Nakakura}
, Yuki \textsc{Nagai}$^{1}$
and Daijiro \textsc{Yoshioka}}

\inst{Department of Basic Science, The University of Tokyo,
 Komaba, Meguro-ku, Tokyo 153-8902 \\
$^{1}$Department of Physics, The University of Tokyo, Hongo, Bunkyo-ku, Tokyo 113-0033}

\kword{graphene, edge state, localized state, electric current}

\begin{document}
\maketitle
Graphene exhibits zero-gap massless Dirac fermions
and zero density of states at $E=0$.\cite{wal}
These particles form localized states called edge states on a finite-width strip with zigzag 
edges at $E=0$.\cite{fujita}
Naively thinking, one may expect that a
current is concentrated at the edge, but 
Mu\~noz-Rojas et~al. noticed that it is homogeneously distributed,\cite{munoz} and Z\^arbo 
and Nikoli\'c\cite{zarbo} numerically obtained a result that 
the current density is maximum at the center of the strip.
We rigorously derive an expression for the current density and clarify the 
reason why the current density of the edge state has a maximum at the center. 

\begin{figure}[h]
\begin{center}
\includegraphics[width=6.cm]{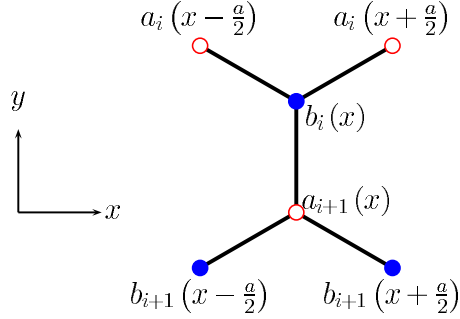}
\end{center}
\caption{(Color online) Definition of wave function.
Part of the strip is shown. In this structure amplitude of the upper (lower) edge state is mainly on the A-sublattice (B-sublattice).
}
\label{fig:label}
\end{figure}

The expectation value of local current density for a state $|\psi\rangle 
$
is defined as
\begin{eqnarray}
\langle \psi |\hat{{j}}_{(x,i),(x',i')}|\psi \rangle  
=\frac{2et}{\hbar }{\rm Im}\left[b_{i'}^{*}(x')a_i(x)\right], \label{kyokusyo2}
\end{eqnarray}
where $\hat{j}_{(x,i),(x',i')}$ is the current density operator of the bond from site $(x,i)$ to 
site $(x',i')$, $a_i(x)$ and $b_{i'}(x')$ are the corresponding amplitudes of the 
wave function at these sites as shown in Fig.~\ref{fig:label},  $e$ is the charge, and
$t$ is the nearest-neighbor hopping parameter.
Because of the translational symmetry along  the $x$-axis,
the amplitudes of the wave function at the A-sublattice and B-sublattice with wave number 
$k$ are 
respectively defined as 
\begin{eqnarray}
& a_i(x)= A_i \exp({\mathrm{i}kx}),\label{ai}\\
& b_{i}(x)= B_{i} \exp({\mathrm{i}kx})\label{bi}.
\end{eqnarray}
The Schr$\ddot{\text{o}}$dinger equations in the tight-binding model 
with nearest-neighbor hopping 
are written as 
\begin{eqnarray}
& KA_i+tA_{i+1}=EB_i ,\label{se1}\\
&tB_i+KB_{i+1}=EA_{i+1}.\label{se2}
\end{eqnarray}
Here, $K=2t\cos \left( ka/2 \right)$, $E$ is the energy, and $a$ is the 
lattice constant.\cite{real}
In the region $2 \pi / 3a < |k| < \pi/a$, a solution with $E\simeq 0$ 
exists, which is the edge state.\cite{fujita}
In this region, 
the wave amplitudes at the A- and B-sublattice are approximately written as 
\begin{eqnarray}
A_i=A_1\left(-2\cos \frac{ka}{2}\right)^{i-1}, \  \   (1\leq i \leq w)\label{asinpuku}\\
B_{i}=\pm A_1\left(-2\cos \frac{ka}{2}\right)^{w-i}.  \  \  (1\leq i 
 \leq w)\label{bsinpuku}
\end{eqnarray}
In these equations $w$ defines the width of the strip. The wave function 
becomes exact in the limit of either $w\rightarrow \infty$ or
$k\rightarrow \pi/a$.

\begin{figure}[h]
\begin{center}
\includegraphics[width=2.5cm]{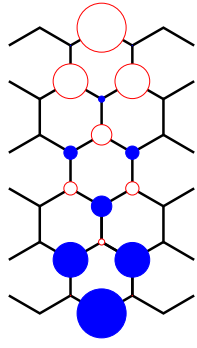}
\end{center}
\caption{(Color online) Wave function of the edge state. The size of each circle shows the amplitude of the wave function at that site, the open red (closed blue) circles reperesent the amplitude of the wave function on the A-sublattice (B-sublattice).
}
\label{f1}
\end{figure}

In this paper, we derive a relation for the local current density
 in the $x$-direction.
 The local current density flowing through the bond between 
 $a_{i}\left(x-a/2\right)$ and $b_{i}(x)$ is 
written as 
\begin{equation}
 j_{i}=\frac{2et}{\hbar } A_{i}B_{i}\sin\frac{ka}{2}.\label{se4}
\end{equation}
Using eqs.(\ref{se1}), (\ref{se2}), and (\ref{se4}), we obtain the relation between 
$j_i$ and $j_{i+1}$ as 
\begin{equation}
j_{i+1}-j_i = \frac{eE}{\hbar } 
  (A_{i+1}^2-B_i^2)\tan\frac{ka}{2}.\label{ji1}
\end{equation}
At $i=1$, namely, at the upper zigzag edge,
\begin{equation}
  j_1=\frac{eE}{\hbar }A_1^2\tan\frac{ka}{2}.\label{se5}
\end{equation}
The $y$-dependence of the local current density is 
calculated from the initial condition [eq.(\ref{se5})] and the recurrence equation [eq.(\ref{ji1})].
In the edge state, where $|E(k)|\leq E(2\pi/3a)=2t \sin [\pi /2(2w+1)]$,\cite{cre} the 
wave function is localized around the edges and is approximately given by 
eqs.(\ref{asinpuku}) and (\ref{bsinpuku}).\cite{waka2}
Notice that the wave function at the A-sublattice is localized 
at the upper edge and that that at the B-sublattice is localized at the lower edge 
as shown in Fig. \ref{f1}.
Therefore, $A_{i+1}^2 - B_i^2 >0$ in the upper half of the strip, and  
the local current density $j_{i}$ increases as $i$ increases.
 At the center of the zigzag strip, $j_i$ is maximum 
since $A_{i+1}^2-B_i^2 = 0$.

When $w\gg 1$ or $k\simeq\pi/a$, the coupling of edges becomes small, the 
wave functions given by eqs.(\ref{asinpuku}) and (\ref{bsinpuku}) become almost 
exact, and $E$ approaches zero.
In such a situation eq.(\ref{ji1}) implies that
$j_{i+1}\simeq j_i$, namely, the current flows almost uniformly. This 
is also implied by eqs.(\ref{asinpuku}) and (\ref{bsinpuku}).
By substituting eqs.(\ref{asinpuku}) and (\ref{bsinpuku}) 
into eq.(\ref{se4}), $j_i$ becomes independent of $i$, and 
is written as 
\begin{equation}
 j_{E\to 0}
=\frac{2et}{\hbar }\left(-2\cos \frac{ka}{2}\right)^{w-1}A_1^2
\sin \frac{ka}{2}. \label{noc} 
\end{equation}
The energy of the edge state in such a situation is obtained from 
eqs.(\ref{se1}) and (\ref{se2}) to be\cite{waka3}
\begin{eqnarray}
 E(k)\simeq \pm t\left(-2\cos \frac{ka}{2}\right)^w 
\left[1-\left(2\cos \frac{ka}{2}\right)^2\right]\label{bunsan}.
\end{eqnarray}
The expression for the current is consistent with that obtained from the group velocity 
\begin{align}
v(k)&=\frac{\partial E(k)}{\partial k} 
\simeq\frac{at}{\hbar}w\left(-2\cos\frac{ka}{2}\right)^{w-1}\nonumber\\
&\times\left[1-\left(2\cos \frac{ka}{2}\right)^2\right]
\sin\frac{ka}{2}, 
\end{align}
at this energy.\cite{energy}
 When $|E|\rightarrow 0$, the magnitude of the current for a state becomes 
small, as implied by eq.(\ref{noc}). However, the conductance is still $2e^2/\hbar$, 
because the smallness of the current is compensated for by the large density of states.

\begin{figure}\centering
\includegraphics[width=5cm]{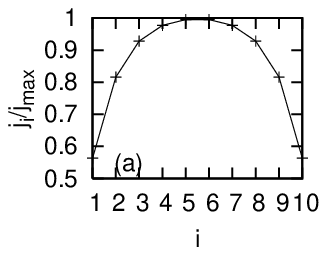}
\includegraphics[width=5cm]{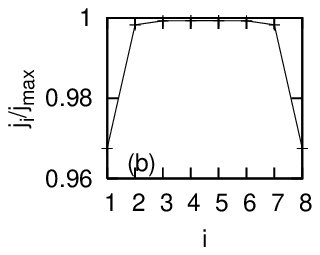} 
\caption{\label{fig:fig1}
(a) Local current density $j_i/j_{\mathrm{max}}$ for $E=0.01t$ and 
   $w=10$, where $j_{\mathrm{max}}$ is the value of $j_i$ at the center. This result is consistent with that of Z$\mathrm{\hat{a}}$rbo 
and Nikoli$\mathrm{\acute{c}}$.\cite{zarbo}
(b) Local current density for $E=10^{-6}t$ and $w=8$.
}
\end{figure}

To verify our analytical discussion, we numerically calculated the current density from eq.(\ref{kyokusyo2}). 
Using Z$\mathrm{\hat{a}}$rbo 
and Nikoli$\mathrm{\acute{c}}$'s parameter $(E = 0.01t, w = 10)$, 
we reproduced their result\cite{zarbo}  as 
shown in Fig. \ref{fig:fig1}(a). 
With decreasing energy, the current flows almost uniformly in 
the $y$-direction as shown in Fig. \ref{fig:fig1}(b).  
These results are consistent with those derived from eq.(\ref{ji1}).

In summary, a finite-width strip with zigzag edges has edge states with 
energy $|E(k)|\leq E(2\pi/3a)$, and electrons are strongly localized 
around the edges. 
However, the current flows almost uniformly at 
the center and decreases towards the edges.   
This behavior is understood by the expression for the local current density 
in the $x$-direction.
This uniform current is realized as a result of three factors.
Firstly, the wave function of the edge states decays exponentially towards the other edge.
Secondly, graphene consists of an A-sublattice and B-sublattice.
Thirdly, the local current density is calculated by the multiplication of 
amplitudes on both sublattices.\cite{munoz}

In this short note we considered the current distribution of an edge state 
specified by wave number $k$.
The interaction between electrons is not 
taken into account.
We remark that the possibility of magnetic ordering of the
edge state has been suggested by several authors.\cite{fujita, son, pis}

\section*{Acknowledgments}
We thank Yusuke Kato and Daisuke Takahashi for helpful discussions.
YN acknowledges support by a Grant-in-Aid for JSPS Fellows (204840).
\appendix


\begin{thebibliography}{99}
\bibitem{real} Since $A_i$ and $B_i$ can be real, we choose them to be real hereafter.
\bibitem{waka3}
The expression for the energy in ref. \citen{waka2}, eq.(3.12), is not correct.
\bibitem{energy}

For a strip of length $la$, 
$A_1^2\simeq 	\{1-\left[2\cos\left(ka/2\right)\right]^2\}/2l$.
 Thus, the total current is $J=wj_{E\to 0}=ev(k)/la$, as expected.
\end{thebibliography}
\end{document}